\documentstyle[twoside,fleqn,espcrc2,epsf]{article}

\newcommand{\be}{\begin{equation}}
\newcommand{\ee}{\end{equation}}
\newcommand{\ba}{\begin{eqnarray}}
\newcommand{\ea}{\end{eqnarray}}
\def\reff#1{(\ref{#1})}

\def\spose#1{\hbox to 0pt{#1\hss}}
\def\ltapprox{\mathrel{\spose{\lower 3pt\hbox{$\mathchar"218$}}
 \raise 2.0pt\hbox{$\mathchar"13C$}}}
\def\gtapprox{\mathrel{\spose{\lower 3pt\hbox{$\mathchar"218$}}
 \raise 2.0pt\hbox{$\mathchar"13E$}}}
\newcommand{\Tr}{\mathop{\rm Tr}\nolimits}
\font\german=eufm10 scaled\magstep1     
\def\germansu{\hbox{\german su}}

\title{Numerical Study of the Gluon Propagator in Lattice Landau Gauge:
       the Three-Dimensional Case}

\author{Attilio Cucchieri\address{Gruppo APE -- Dipartimento di Fisica,
        Universit\`a di Roma ``Tor Vergata'', 
        and INFN -- Sezione di Roma 2, Via della Ricerca Scientifica 1,
        I-00133 Roma, ITALY}\thanks{Presented in the Poster session.
                Address after September 30, 1998: 
                Fakult\"at f\"ur Physik, Universit\"at Bielefeld,
                Universit\"atsstrasse, D-33615 Bielefeld, GERMANY.}}
       
\begin{document}

\begin{abstract}
We study the infrared behavior of the gluon propagator in lattice
Landau gauge, for pure $SU(2)$ lattice gauge theory in a
three-dimensional lattice. Simulations are done for nine different values
of the coupling $\beta$, from $\beta = 0$ (strong coupling) to
$\beta = 6.0$ (in the weak-coupling region). In the limit of large
lattice volumes, we observe in all cases a gluon propagator decreasing
as the momentum goes to zero.
\end{abstract}

\maketitle

\section{INTRODUCTION}

The infrared behavior of the gluon propagator in lattice Landau gauge has been
the subject of several numerical studies (see
\cite{Athesis,Attilio}
and references therein).
In fact, although
this propagator is a non-gauge-invariant quantity, the study of its
infrared behavior provides a powerful tool for increasing our understanding of
QCD, and for gaining insight into the physics of confinement in non-abelian gauge
theories.
 
On the lattice, the Landau gauge condition is imposed
by finding a gauge transformation which brings the functional ${\cal E}_{U}[ g ]$,
defined in eq.\ \reff{eq:Etomin} below, to a minimum. A lattice configuration which
satisfies this minimizing condition belongs to the region $\Omega$ of transverse
configurations, for which the Faddeev-Popov operator is nonnegative
\cite{DZ,Z1}.
This region is delimited by the so-called first Gribov horizon, defined as
the set of configurations for which the smallest non-trivial eigenvalue
of the Faddeev-Popov operator is zero.

The restriction of the path integral, which defines the partition function,
to the region $\Omega$ implies a {\em rigorous} inequality
\cite{DZ,Z1}
for the Fourier components of the gluon field $ A $. From this inequality, which is a
consequence only of the positiveness of the Faddeev-Popov operator, it
follows that the region $\Omega$ is bounded by a certain
ellipsoid $\Theta$. This bound implies proximity of the first Gribov horizon
in infrared directions, and consequent suppression of the low-momentum
components of the gauge field, a result already noted by Gribov in Ref.\
\cite{Gr}.
This bound also causes a strong suppression of the gluon propagator
in the infrared limit (i.e.\ for momentum $p \to 0$). More precisely, Zwanziger proved
\cite{Z1,Z2}
that, in the infinite-volume limit, the gluon propagator is less singular than $p^{-2}$
in the infrared limit and that, very likely, it vanishes as $p^{2}$.
A gluon propagator vanishing as $p^{2}$ in the infrared limit was also
found (under certain hypotheses) by Gribov
\cite{Gr}. Finally, a gluon propagator vanishing in the infrared limit has
been recently obtained as an approximate solution of the gluon
Dyson-Schwinger equation
\cite{smekal}.

Here we study the infrared behavior of the gluon propagator in the
three-dimensional case (preliminary results have been reported in Ref.\
\cite{lat97}).
Let us recall that
nonabelian gauge theories in three dimensions are similar to
their four-dimensional counterparts, and results obtained in the
three-dimensional case can teach us something about the more
realistic four-dimensional theories.
Of course, the advantage of using a three-dimensional lattice is the possibility
of simulating lattice sizes larger than those used in the four-dimensional case.
This is particularly important in the study of the gluon propagator: in fact,
Zwanziger's prediction
\cite{Z1,Z2}
of an infrared-suppressed gluon propagator is valid only
in the infinite-volume limit.


\section{METHODOLOGY}
\label{sec:propag}
 
We consider a standard Wilson action for $SU(2)$ lattice gauge theory
in $3$ dimensions, with periodic boundary conditions. (Details of
notation and numerical simulations will be given in
\cite{future}.)
The gauge field,
which belongs to the $\germansu\, (2)$ Lie algebra, is defined as
$ A_{\mu}(x) \equiv 1/2 \left[\,U_{\mu}(x)\,-\,U_{\mu}^{\dagger}(x)
\,\right] $, where $U_{\mu}(x) \in SU(2)$ are link variables.
We also define $ A_{\mu}^{a}(x) \equiv \Tr \left[ \, A_{\mu}(x) \,
\sigma^{a} \, \right] / (2 i) $, where $\sigma^{a}$ is a Pauli matrix.

In order to fix the lattice Landau gauge we look for a local minimum
of the functional
\be
{\cal E}_{U}[ g ]\,\equiv\,1\,-
       \frac{1}{3\,V}\,\sum_{\mu = 1}^{3}\,
  \sum_{x}\,\frac{\Tr}{2}\,U_{\mu}^{(g)}(x)
\;\mbox{,}
\label{eq:Etomin}
\ee
where $ V $ is the lattice volume, and
$ U_{\mu}^{(g)}(x) \equiv g(x)\,U_{\mu}(x)\,g^{\dagger}(x + e_{\mu}) $.
[Here $\,g(x)\in SU(2)\,$ are site variables.]

Finally, the lattice gluon propagator in momentum space is defined as
\ba
D(0) \! \!\!& \equiv& \! \!\! \frac{1}{9 V} \sum_{\mu\mbox{,}\,a}\,\langle\,
  \left[\,\sum_{x}\,A_{\mu}^{a}(x)\,\right]^{2} \rangle
\label{eq:D0def} \\
D(k) \! \!\! & \equiv & \!  \!\!\frac{1}{6 V} \sum_{\mu\mbox{,}\,a}\,\langle\,
\left\{\left[\,\sum_{x}\,A_{\mu}^{a}(x)\,
\cos{( 2 \pi k \cdot x )}\,\right]^{2} \right. \nonumber \\
                                        \!\!\! & &  \!\!\! \left.
 + \left[\,\sum_{x}\,A_{\mu}^{a}(x)\,
\sin{( 2 \pi k \cdot x )}\,\right]^{2}
\right\} \,\rangle
\;\mbox{.}
\label{eq:Dkdef}
\ea

 
\section{RESULTS}
\label{sec:infra}

In Figures \ref{fig:gluo1} and \ref{fig:gluo2} we plot the data for the
gluon propagator 
as a function of the square of the lattice momentum
$ p^{2}(k)
\equiv 4 \sum_{\mu = 1}^{3} \sin^{2}{\left( \pi \,k_{\mu} \right)} $,
for different lattice volumes $V$ and couplings $\beta$.
Our data confirm previous results
\cite{Athesis,Attilio}
obtained in the strong-coupling regime for the four-dimensional case: the
gluon propagator is decreasing as $p$ decreases, provided that $p^{2}(k)$
is smaller than a turn-over value $p^{2}_{to}$. Clearly $p^{2}_{to}$ is
$\beta$- and volume-dependent. Also, as in four dimensions, the lattice
size at which this behavior for the gluon propagator starts to be observed
increases with the coupling. In particular, in the strong-coupling regime,
this propagator is clearly decreasing as $p^{2}(k)$ goes to zero,
even for relatively small lattice volumes (see the case $\beta = 2.8\, $
in Figure \ref{fig:gluo1}).
On the contrary, for $\beta \geq 3.4$, this propagator is increasing
(monotonically) in the infrared limit for $V = 16^3$, while it is decreasing
(see Figures \ref{fig:gluo1} and
\ref{fig:gluo2}) for the largest lattice volume considered.
 
Let us also notice that, at high momenta, there are very small finite-size
effects, at all values of $\beta$. The situation is completely different in
the small-momenta sector, as already stressed above. In particular, the value
$D(0)$ of the gluon propagator at zero momentum decreases monotonically as
the lattice volume increases (see for example the case $\beta = 5.0$ in Figure
\ref{fig:gluo2}). These results suggest a finite value for $D(0)$ in the
infinite-volume limit, but it is not clear whether this value would be zero or a
strictly positive constant. Therefore, the possibility of a zero value for
$D(0)$ in the infinite-volume limit is not ruled out.

\section{CONCLUSIONS}

%
As said above, our data in the strong-coupling regime are in qualitative agreement
with the results obtained in the four-dimensional case (see Figure 1 in Ref.\
\cite{Attilio}).
This strongly suggests to us that a similar analogy will hold
--- in the limit of large lattice volumes --- also for couplings $\beta$
in the scaling region, leading to an infrared-suppressed gluon
propagator in four dimensions.

\begin{figure}[hbt]
\begin{center}
\vspace*{0cm}
\epsfxsize=0.41\textwidth
\protect\hspace*{0.4cm}
\leavevmode\epsffile{Plot_Vgluon28_axis.ps}
\vspace*{2.0cm}
\epsfxsize=0.41\textwidth
\protect\hspace*{0.6cm}
\leavevmode\epsffile{Plot_Vgluon34_axis.ps}
\end{center}
\vspace*{0.95cm}
\caption{~Plot of the gluon propagator $D(k)$
as a function of the square of the lattice momentum
$p^{2}(k)$ for lattice volumes $V = 16^{3}$
($\Box$) and $V = 32^{3}$ ($\ast$), with $k = (0\mbox{,}\, 0\mbox{,}\,
k_{t})$, at: ({\bf a}) $\beta = 2.8$ and ({\bf b}) $\beta = 3.4$.
Error bars are one standard deviation.
}
\label{fig:gluo1}
\end{figure}
 
\begin{figure}[hbt]
\begin{center}
\vspace*{0cm}
\epsfxsize=0.41\textwidth
\protect\hspace*{0.1cm}
\leavevmode\epsffile{Plot_Vgluon42_axis.ps}
\vspace*{2.0cm}
\epsfxsize=0.41\textwidth
\protect\hspace*{0.3cm}
\leavevmode\epsffile{Plot_Vgluon50_axis.ps}
\end{center}
\vspace*{0.8cm}
\caption{~Plot of the gluon propagator $D(k)$
as a function of the square of the lattice momentum
$p^{2}(k)$ for lattice volumes $V = 16^{3}$
($\Box$), $V = 16^{2}$x$32$ ($+$), $V = 32^{3}$ ($\ast$), $V = 32^{2}$x$64$
($\circ$), and $V = 64^{3}$ ($\Diamond$), with $k = (0\mbox{,}\, 0\mbox{,}\,
k_{t})$, at: ({\bf a}) $\beta = 4.2$ and ({\bf b}) $\beta = 5.0$.
Error bars are one standard deviation.
}
\label{fig:gluo2}
\end{figure}


\end{document}